%% file: Main_text.tex
\documentclass[preprint,12pt,authoryear]{elsarticle}

\usepackage[tableposition=top]{caption}

\usepackage[most]{tcolorbox}

\usepackage{xcolor}
\usepackage{colortbl}
\usepackage{balance}
\usepackage{enumitem}

\newcommand{\sectopic}[1]{\vspace{0.2em}\par\noindent{\textit{\bfseries #1}}}

\definecolor{Always}{RGB}{122, 197, 205}
\definecolor{Never}{RGB}{184, 134, 11}

\usepackage{amssymb}
\usepackage{rotating}
\usepackage{booktabs}

 \usepackage{amsthm}

\usepackage{lineno}

\usepackage{todonotes}

\journal{Information and Software Technology}

\begin{document}

\begin{frontmatter}



\title{Who Uses Personas in Requirements Engineering: \\The Practitioners' Perspective}


\author [label1] {Yi Wang} \ead[label1]{xve@deakin.edu.au}

\author [label2] {Chetan Arora} \ead[label1]{chetan.arora@monash.edu}

\author [label1] {Xiao Liu} \ead[label1]{xiao.liu@deakin.edu.au}

\author [label1] {Thuong Hoang} \ead[label1]{thuong.hoang@deakin.edu.au}

\author [label2] {Vasudha Malhotra} \ead[label1]{vasu.malhotra@monash.edu}

\author [label1] {Ben Cheng} \ead[label1]{chengye@deakin.edu.au}

\author [label2] {John Grundy} \ead[label2]{john.grundy@monash.edu}

\affiliation[label1]{organization={Deakin University },
            city={Geelong},
            state={VIC},
            country={Australia}}
            
            \affiliation[label2]{organization={Monash University},
            city={Melbourne},
            state={VIC},
            country={Australia}}

\begin{abstract}
Personas are commonly used in software projects to gain a better understanding of end-users' needs. However, there is a limited understanding of their usage and effectiveness in practice. This paper presents the results of a two-step investigation, comprising interviews with 26 software developers, UI/UX designers, business analysts and product managers and a survey of 203 practitioners, aimed at shedding light on the current practices, methods and challenges of using personas in software development. Our findings reveal variations in the frequency and effectiveness of personas across different software projects and IT companies, the challenges practitioners face when using personas and the reasons for not using them at all. Furthermore, we investigate the coverage of human aspects in personas, often assumed to be a key feature of persona descriptions. Contrary to the general perception, our study shows that human aspects are often ignored for various reasons in personas or requirements engineering in general. Our study provides actionable insights for practitioners to overcome challenges in using personas during requirements engineering stages, and we identify areas for future research.
\end{abstract}




\end{frontmatter}


\input{Section/Introduction}

\input{Section/Relatedwork}

\input{Section/Methdology}

\input{Section/Results}

\input{Section/Discussion}
\input{Section/Limitation}

\input{Section/Conclusion}

\bibliographystyle{elsarticle-harv} 
\bibliography{reference.bib}


\end{document}

%% file: Section/Introduction.tex
\section{Introduction}

Requirements engineering (RE) focuses, among others, on understanding the needs and goals of the users who will interact with the system under development~\citep{pohl2010requirements}. One effective way to achieve this is through the use of personas~\citep{Salminen22}. Personas are fictional (yet specific and concrete) characters that represent different user types, and they help to create a common understanding of the users' demographics, needs, behaviors, motivations, and pain points~\citep{Schneidewind12,dupree2018case,Jianing22,Salminen22}. By using personas, requirements engineers can design systems that meet the needs of the intended users and ensure that the system is user-friendly and effective~\citep{blomquist02, adlin01, Anj18}. 


Personas have been widely adopted in numerous fields, such as marketing~\citep{revella2015buyer}, product design~\citep{long2009real}, education design~\citep{arora2023persona}, and user experience design~\citep{matthews2012designers}. In RE, personas are anecdotally known to be widely used~\citep{Aoyama05,Schneidewind12}. In the early stages of RE, personas are known to be typically created based on research and analysis of user data, such as surveys, interviews, and usability tests of previous versions of the software~\citep{Salminen22, pruitt2003}. Despite the anecdotal evidence on the widespread use of personas in software development, there is limited research on how practitioners use them in RE. For instance, to what extent are personas used in practice, the kinds of projects where personas are more likely to be used, or the challenges associated with implementing personas in practice? There has not been much research to answer such questions.

To address this gap, this paper presents the results of a survey and interviews with practitioners to explore the status quo of persona usage in the industry.  Our main objective of the survey and interviews was to investigate the use of personas in industry, specifically including (1) the use of personas in software projects for customers, businesses, and government in different-sized companies, e.g., large corporations, small-to-medium enterprises (SMEs) and startups; (2) the key job roles involved in using and creating personas in the industry; (3) the methods utilized by different roles and companies to implement personas, including approaches for identifying the key `human aspects' of users~\citep{Hidellaarachchi22} and providing feedback on current methods or standards, and (4) the challenges associated by practitioners with personas.


As mentioned above, one key factor (among others) we were interested in personas was the coverage of human aspects, such as personality, gender, motivation, culture, and physical or mental issues~\citep{Grundy20,fazzini2022characterizing}. Human aspects have recently received increasing attention in RE research~\citep{wang21,Hidellaarachchi22,Hidellaarachchi22ACM,ahmad2023requirements,Arora:23} and have become an integral part of the software development process, as developers aim to develop systems that are more responsive to user needs and better align with the user goals and values. Given that personas, by their very definition, emphasize the individual attributes (or human aspects) of users, we felt it was imperative to include a section in our survey and interviews specifically examining the extent to which these aspects are represented in personas.

We surveyed 203 practitioners and interviewed 26 practitioners to understand the usage (or not) of personas in their context. The survey included software developers, UI/UX designers, business analysts, project managers, and user researchers. User researcher was a slightly unconventional role that we came across in our survey. The role is responsible for understanding user behavior, needs, and motivations to inform the design and development of products and digital services. In our interviews, we had similar roles and a CEO of a startup for an e-learning system. We had good coverage of participants from large corporations, SMEs and startups, and people working in different application domains, such as finance, education, cloud services, and virtual reality (VR).

Our study revealed that the usage of user personas varies significantly among different types of software projects and different-sized IT companies. For instance, personas are less commonly used in startups, government software projects, or in VR and gaming projects. Moreover, various IT companies and practitioners tend to have diverse preferences for methods to collect and analyze user data and generate personas. Some companies deploy dedicated data-driven platforms for collecting and maintaining user information for personas, whereas some companies (including startups) favor traditional RE methods to analyze user data and capture user requirements. Moreover, we observed that the human-aspect-information within personas can differ based on distinct application contexts. Notably, certain user groups, such as those with specific physical needs, often go unnoticed. Finally, based on our analysis, we identify some current challenges related to using personas in industry and propose future research directions. In summary, the paper makes the following contributions:

\begin{itemize}
\item To the best of our knowledge, this is the first study to provide empirical evidence on the extent to which personas are used in RE and the factors influencing their adoption. Our empirical evidence is based on a survey with more than 200 practitioners and interviews with 26 practitioners.

\item We report on the types of projects and organizations in which personas are most commonly used and the factors that drive their adoption in these contexts.

\item We delineate the key challenges associated with persona adoption in the industry. 

\item We identify gaps and recommendations for future research on the use of personas in RE.
\end{itemize}

%% file: Section/Relatedwork.tex
\section{Related Work}\label{RW}



\textcolor{black}{Personas are a critical approach for UI/UX design and development teams (e.g., large development teams) to gain a better understanding diverse users \citep{matthews2012designers, cooper2014face, goodwin2011designing}. Prior work has noted that personas are used more widely in software development than in other areas such as healthcare and education \citep{Salminen22}.} 
Billestrup et al.~\cite{Billestrup14} investigated the use of personas in software development and found that personas can facilitate collaboration between designers and developers not working closely together. However, using personas may also bring some risks, such as when personas describe the wrong user groups. To address the limitations of personas, \citet{Watanabe20} proposed using data-driven methods to generate personas to support business-to-business software development companies \citep{salminen2021s}. \textcolor{black}{\citet{zhang2023personagen} demonstrated a generating persona tool based on the GPT-4 model and knowledge graph. }\citet{Pruitt03} pioneered the guidelines for using personas as an essential medium for communication among development teams. Additionally, \citet{Anvari17} suggested that personas should include human factors to represent the target users of software applications better~\citep{Dow06, Zhu19}.

Personas are frequently used in different RE phases, such as elicitation and validation~\citep{dupree2018case,Jianing22,Salminen22}. For instance, \citet{Silvia12} have argued that personas can enhance the understanding of software product goals, behavior, and the characteristics of users and their interactions with the systems. Persona's content must combine the various phases of RE and design processes with user insights \citep{Salminen22, Anvari15, Schneidewind12, grudin2002}. In particular, different personas can be allocated to different designers and software developers in software projects. Software developers can also use ad-hoc or imaginary personas to analyze potential requirements in RE~\citep{cleland13}. \textcolor{black}{Although the literature on the use of personas in RE is limited, a recent study highlighted that personas have notable shortcomings in RE activities, particularly in the analysis, specification, and validation of requirements~\citep{KAROLITA23perosna}. }Many methods are combined with personas to analyze requirements and guide design and interface development, such as scenarios, semi-structured interviews, prototyping, and storyboards~\citep{Pruitt03, Anvari17, haan15, arvola18}. For example, prior research has proposed a method that automatically generates 3D avatars based on persona data by observing user behaviors in the virtual environment~\citep{wang21}. \citet{blanco14} have found that scenarios and personas can help improve shared understanding of functional requirements, thereby reducing ambiguities in the requirements. Scenarios can also help to complement the contexts of personas and provide a clearer description of user requirements in RE~\citep{Aoyama07, Aoyama05}. \textcolor{black}{\citet{Salminen22} proposed a survey instrument (Persona Readiness Scale) to measure organizational readiness for personas. This approach can help to improve compatibility between the organizations and personas.}

\citet{Grundy20} pointed out that human-centric issues are incorporated into the critical SE challenges. Some researchers have investigated important human factors such as personality, emotions, motivation, culture, and so on~\citep{luiz10,colomo19,franca14, hina12}. However, due to a lack of knowledge of the design and evaluation of these human factors, software engineers can not incorporate human factors issues into the development process~\citep{Jianing22, Winter18}. Thus, some persona prototype tools have been designed to help developers address challenges in their software projects~\citep{Jianing22, Salminen22}. \citet{Jianing22} demonstrated how persona prototype tools are used in industry. More than half of the participants agreed that the persona prototype tool helps discover potential issues such as accessibility needs~\citep{Almeida21}.

All these existing works cover several important factors related to persona creation, information captured in personas, and the importance of personas in RE and software development. However, no study (to the best of our knowledge) has covered the practitioners' perspective of personas and the status quo of persona usage in the industry, which our study attempts to do with more than 203 practitioners in our survey and 26 participants in our interviews.

%% file: Section/Methdology.tex
\section{Methodology}\label{mdy}
Our study aims to understand the practitioners' perspectives on the usage of personas in different types of companies and industries for different types of software projects. We base our study on a survey and semi-structured interviews to collect different types of self-reported data. \textcolor{black}{The Human Ethics Advisory Group at Deakin University (Faculty of Science, Engineering and Built Environment) reviewed and approved our study (reference: SEBE-2022-62).} \textcolor{black}{Our material used for conducting interviews and surveys is publicly available in both English and Chinese languages~\citep{Whouse}. }

\subsection{Research Questions}\label{rqs}
\textcolor{black}{Our study was structured around three research questions (RQs):}

\textbf{RQ1. How common is the use of personas across different software development projects?}
RQ1 aims to investigate the current state of persona usage in the industry for software development. We analyze the type of companies that employ personas (e.g., big corporations, small and medium-sized enterprises (SMEs), and startups), the type of software projects at hand (e.g., the software as part of corporate projects, government projects, and developed directly for end users), and the roles that use personas.

\textbf{RQ2. What are the current practices for building personas?} RQ2 aims to understand the current practices around capturing the users' information in personas. To this end, among others, RQ2 delves into the inclusion of human aspects — delineated as human attributes such as age, gender, cultural background, and physical and mental abilities — in the persona construct.

\textbf{RQ3. What are the main challenges associated with persona usage in software projects?} RQ3 aims to explore the challenges that practitioners associate with the implementation of personas in software projects, to understand the reasons for their limited adoption, and ultimately assess the feasibility of addressing some of these challenges.

To answer these RQs, we adopted a sequential exploratory strategy to comprehensively investigate the use and construction of personas in software development~\citep{creswell2017research}. Our approach began with in-depth, qualitative interviews with 26 practitioners spanning diverse roles and organizations. The interviews offered valuable preliminary insights, helping us understand the nuances, key themes, and patterns around persona usage.  Informed by these findings, we designed and administered a survey to 203 practitioners to quantitatively examine our findings on a broader scale. Below, we discuss the design of our interviews and survey studies.



\input{Section/tab1}

\subsection{Participants Recruitment}
In our study, we aimed at recruiting a diverse cohort of participants. \textcolor{black}{ To ensure the relevance of our findings, we mandated that participants have at least three years of work experience and are familiar with personas. Additionally, for survey respondents we also wanted to ensure that they held positions in IT/SE domains or were associated with companies operating within the IT/SE sectors (some responses were filtered out during analysis).} We advertised our study to various social media, such as LinkedIn \citep{LinkedIn}, WeChatp~\cite{WeChat}, and Twitterp~\cite{Twitter}. We also recruited some interviewees via our professional networks. For the online survey, we distributed the online survey through Web forums (e.g., GitHub \citep{GitHub} and Reddit \citep{Reddit}), and social media (e.g., LinkedIn \citep{LinkedIn}, Twitter \citep{Twitter} Discord-related Groups~\citep{discord} and WeChat~\citep{WeChat}). After the interviews, we also sent our survey to interviewees and asked them to fill out and distribute it to related departments within their companies. We provided each interviewee with an incentive of 20 USD Amazon e-gift cards, 30 AUD Amazon e-gift cards, and 150 RMB red envelopes using WeChat. Each survey participant received an incentive of 3 USD, 5 AUD or 20 RMB, online gift cards or WeChat red envelopes.

\subsection{Semi-structured Interviews Design}~\label{subsec:interviewsDesign}
As the first step of our study, we conducted a series of one-to-one semi-structured interviews with professional software developers, UI/UX designers, IT project managers, business analysts, and user researchers. \textcolor{black}{The interviews were conducted via Zoom~\citep{zoom21} and VooV}~\citep{voov21}. The semi-structured interviews were based on an interview guide that included questions on the advantages and disadvantages of using personas, human aspects, special user groups in persona creation, and usage of use personas in the software development lifecycle~\citep{Bacchelli13}. As practitioners had various roles and work experiences, we iteratively refined our interview guidelines after each interview. The final interview was organized into the following main stages:

\sectopic{(1) Introduction and background questions.} In this initial phase, we briefly outlined our study, introducing RE, human aspects, and personas. Before commencing the interview, we reiterated key points from the consent form, ensured it was signed by the participants, and informed them about the audio recording. We emphasized that participants could skip questions they found uncomfortable and had the freedom to halt the interview whenever they wished. Following this, we initiated the audio recording and gathered demographic information from participants, such as their years of work experience, types of previous software projects, and familiarity with using personas. Discussing the background acted as an icebreaker and provided us with additional context to analyze data.

\textbf{(2) General persona questions.} In this section, we asked participants general questions about the usage of personas in software projects. We asked about the advantages and disadvantages of using personas to analyze software requirements, how personas can be used to derive user requirements, and how personas can be covered in the software development lifecycle. We also asked more detailed questions based on the participants' work experience. \textcolor{black}{However, some participants reported having limited experience using personas in their current role, but they all understood them from previous software projects.}

\textbf{(3) Human aspects and special user groups questions.} In this section, we asked participants about the inclusion of human aspects and special user groups in personas in software projects. Firstly, we asked if participants considered human characteristics and special user groups in personas. For those who confirmed they did, we asked self-reported questions to determine whether human aspects or special user group personas helped improve requirements quality, user satisfaction, success rate, etc. We also asked participants to identify which human aspects or special user groups are important in software projects. For those not using these components in their personas, we introduced example human aspects, such as personality and motivation, and special user groups, like low-vision users, inviting their perspectives on the potential merits of these inclusions.

\textbf{(4) Personas usage questions.} In this interview section, we aimed to gather information on how personas are commonly used in different software projects, including business, customer, and government projects. We refer to Table \ref{tab:tab1} for the details of the interviewees' experience. For those with limited experience using personas for user requirements, we asked about the requirements analysis methods they used in various software projects. Additionally, we asked participants to discuss current methods' limitations and benefits. Based on their responses, we further probed their thoughts on how personas could address some of the identified limitations and provide additional comments.

\textbf{(5) Wrap-up.} We concluded the interviews with general questions, such as ``Are there any additional comments or suggestions about using personas for user requirements in software projects?'' and ``Do you have any questions for us?'' We rounded off by expressing our appreciation for the interviewees' invaluable inputs.

\subsection{Survey Design}\label{survey}
In the second phase of our study, we rolled out a survey targeting practitioners to answer our RQs. \textcolor{black}{Our objective was to validate and expand on the findings garnered from the semi-structured interviews and further answer our RQs more comprehensively (see §\ref{rqs}). }We created the survey in two languages - in English using Qualtrics~\citep{qu21} and in Chinese using Tencent Questionnaire~\citep{te21} and Wenjuanxing~\citep{wjx21}.
\textcolor{black}{The survey had 37 closed-ended questions in total, which included six single-choice, four multiple choice, seven items on a 5-point Likert scale, and three items on a 5-point matrix rating scale questions (ranging from `strongly disagree' to `strongly agree'). To ensure the quality of the answers, all survey questions were not required fields (see §\ref{analysis}). The survey was structured into five blocks of questions. In the following, we detail the design of the surveys and outline the participant's progress through each block of questions.}



\textbf{(1) Introduction and consent page.}  \textcolor{black}{The survey began with an introduction, followed by sections detailing its purpose and procedure. The inclusion criteria included a reminder advising participants unaware of personas not to complete the online survey. The survey also collected personal information and provided assurances regarding data confidentiality and contact information should participants have any questions.} We also informed participants that the study adheres to the low-risk ethical guidelines and requested their consent to use their data. Participants were also informed that they could withdraw from the study if they felt uncomfortable.

\textbf{(2) Demographics.} This segment solicited demographic details from participants, encompassing aspects like their job roles, years of professional experience, and the nature of their organizations. \textcolor{black}{Demographic questions were included to help us identify respondents' information and understand the attitudes of different types of participants towards personas \citep{falessi2018}.}


\textbf{(3) General questions about personas.} \textcolor{black}{In this survey section, we asked participants general questions about the feedback of using personas in software projects. For example, these questions included assessment factors such as costs, time, and value.}



\textbf{(4) Personas usage and creation-related questions.} In this section, we explored participants' usage of personas in various domains, including business, government and customer projects, and virtual reality projects. We also inquired about the methods used to collect and analyze data for persona creation. We also asked participants how personas are used in different types of companies. 

\textbf{(5) Human aspects related questions.} In the final section, we asked participants questions to collect information about three main types of human aspects: individual-related, technical-related, and team-related. We followed the design proposed by \citet{Hidellaarachchi22} to create questions about these human aspects. Our objective was to explore the value of using personas for capturing human aspects in software requirements.

\begin{figure*}[!t]
  \centering
  \includegraphics[width=\linewidth]{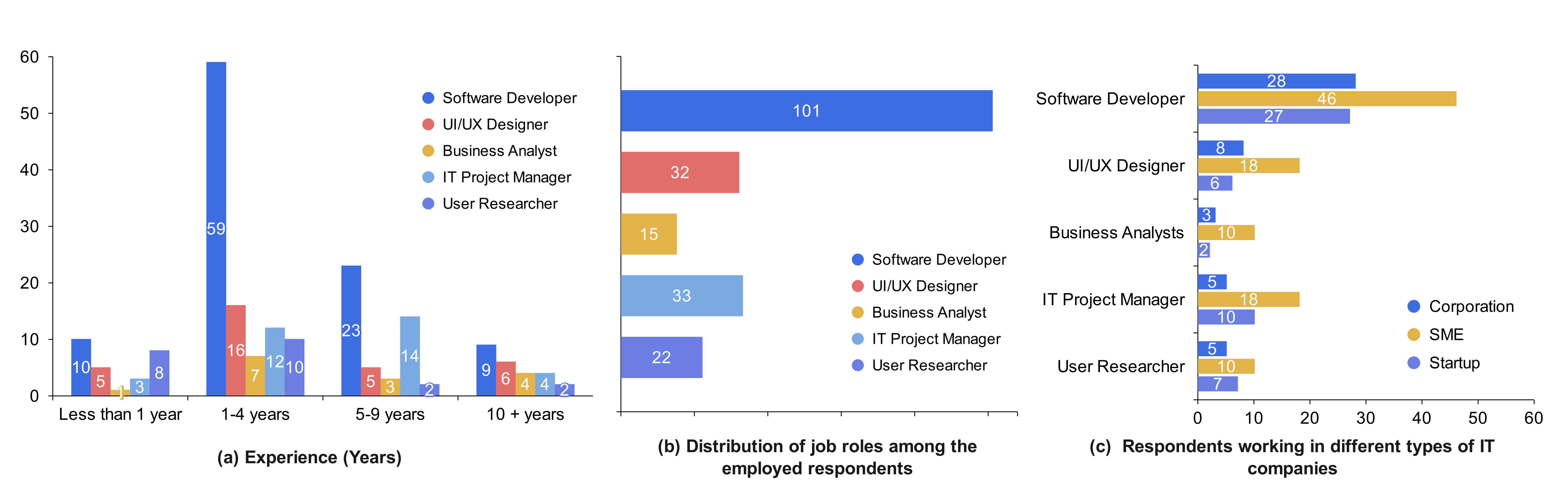}
  \caption{Survey Respondents' Demographics.}
   \label{fig:fig1}
\end{figure*}
    
     

 %


    
     


\subsection{Data Analysis of the Survey and Interviews}\label{analysis}


To analyze the interview data, we used the iFLYTEK professional automated transcription service to transcribe the interview recordings into transcripts \citep{iFLYTEK}. We also translated the Chinese transcript into English, jointly completed by the first author and two external researcher assistants. \textcolor{black}{To evaluate the accuracy of the data, the first author and two co-authors performed an initial accuracy check on the auto-transcriptions. They were proficient in both English and Chinese. This quality check aimed to rectify errors emerging from the automated transcription and organize the structure of the transcriptions. Then, the first author and two authors reviewed the transcripts and conducted a thematic coding analysis on the entire dataset using NVivo~\citep{clarke2015, nv21}. }During the thematic coding process, we screened out sentences that were not related to our study. We used the card-sorting approach to analyze and categorize the interview. Specifically, the first author, two co-authors, and two external researcher assistants read and coded the interview data into cards. To ensure no repeat card themes, the authors merged the cards with the same words, themes, or meanings. The other authors then discussed and reviewed the final card themes. In this discussion, we evaluated the themes, reduced some potential misinterpretations of the participants' answers, and reviewed the card sort output. After reviewing the cards through discussions, \textcolor{black}{we reached the final four main card themes, including 23 card sub-themes based on three research questions and participants' demographics.}

To analyze the online survey data, we first combined the results of the Chinese and English questionnaires. \textcolor{black}{We had initial 323 responses. We excluded any incomplete, \textcolor{black}{unfamiliar with personas,} and nearly identical responses. After this filtering step, we had a total of 203 responses. We categorized the survey results based on demographic information, such as participants' roles, years of experience, company size, and types of companies. We then imported the survey data into the RStudio and analyzed the survey data based on the statistical description \citep{RStudio}. The online survey questions were based on our research questions and interview findings.  }


%% file: Section/tab1.tex
\begin{sidewaystable}

\begin{center}
\begin{minipage}{\textheight}
  \caption{Profile of the interview participants (\textit{N=26})}
  \label{tab:tab1.1}
  \begin{tabular*}{\textheight}{llp{4cm}p{4.2cm}p{1.6cm}c}
   
    \hline
  \textbf{ID} & \textbf{Company Size} &  \textbf{Job Roles} & \textbf{Project Types}& \textbf{Exp. (Years)} & \textbf{Persona Usage}\\ \hline  

    P1 & Corporation & UI/UX Designer&  Government: Banking App, Financial System & 13 & \cellcolor{Always!90} Always\\ \hline
   
    P2 & SME & UI/UX Designer  & A variety of short-term software projects for businesses and customers & 4 & \cellcolor{Always!50} Often\\ \hline

    P3 & Startup & CEO & Customer: Education, E-Learning  & 14 & \cellcolor{Never!90} Never\\ \hline
    
    P4 & Startup  & Software Developer &  Customer: VR, AR, \& Mobile Games & 12 & \cellcolor{Never!50} Rarely\\ \hline
    
    P5 & Corporation & UI/UX Designer &  Customer: Healthcare \& E-Shopping Apps & 6 & \cellcolor{Always!50} Often\\ \hline
    
    P6 & Corporation & Senior Software Engineer&  Customer: All-in-one Mobile Travel App  & 17 & \cellcolor{Always!90} Always\\ \hline
    
    P7 & Corporation & Software Developer &  Customer: All-in-one Mobile Travel App  & 7 & \cellcolor{Never!90} Never\\ \hline
    
    P8 & Corporation & IT Project Manager &  Business: CRM System  & 6 & \cellcolor{Always!90} Always\\ \hline
     
    P9 & Corporation & Software Developer  &  Business: CRM System & 6 & \cellcolor{Never!50} Rarely\\ \hline
     
    P10 & SME & Software Developer  &  Customer: VR Social Media, Desktop Game & 16 & \cellcolor{Never!50} Rarely\\ \hline
    
    P11 & Startup & UI/UX Designer  &  Business: CRM System  & 7 & \cellcolor{Never!90} Never\\ \hline
    
    P12 & SME & IT Project Manager  &  Customer: Education App & 6 & \cellcolor{Always!90} Always\\ \hline
    
    P13 & Corporation & Business Analyst &  Customer: Social Media App & 5 & \cellcolor{Always!90} Always\\ \hline

\end{tabular*}

\end{minipage}
\end{center}

\end{sidewaystable}

\begin{sidewaystable}

\begin{center}
\begin{minipage}{\textheight}
  \caption{(\ref{tab:tab1.1} following. )Profile of the interview participants (\textit{N=26})}
  \label{tab:tab1}
  \begin{tabular*}{\textheight}{llp{4cm}p{4.2cm}p{1.6cm}c}
   
    \hline
  \textbf{ID} & \textbf{Company Size} &  \textbf{Job Roles} & \textbf{Project Types}& \textbf{Exp. (Years)} & \textbf{Persona Usage}\\ \hline

    P14 & Corporation & Business Analyst &  Customer: Social Media App & 6 & \cellcolor{Always!90} Always\\ \hline

    P15 & Corporation & IT Project Manager  &  Customer: Education App & 6 & \cellcolor{Always!90} Always\\ \hline
    
    P16 & Corporation & Senior Software Engineer &  Business: CRM System & 12 & \cellcolor{Never!90} Never\\ \hline

    P17 & Startup & IT Project Manager  &  Customer: Online Survey Software & 5 & \cellcolor{Always!90} Always\\ \hline

    P18 & Startup & UI/UX Designer &  Customer: Healthcare \& Fitness Apps& 6 & \cellcolor{Always!90} Always\\ \hline

    P19 & SME & Software Developer  & Business: Cloud Server Support & 3  & \cellcolor{Never!90} Never\\ \hline    
    
    P20 & Corporation & Senior Software Engineer  &  Business: CRM System & 8 & \cellcolor{Never!50} Rarely\\ \hline

    P21 & Corporation & Software Architect &  Business: CRM System & 4 & \cellcolor{Never!90} Never\\ \hline

    P22 & SME  & Software Developer  & Customer: Learning Management System & 3  & \cellcolor{Never!90} Never\\ \hline

    P23 & Startup & Software Developer & Customer: Accounting Software & 3 & \cellcolor{Never!90} Never\\ \hline
      
    P24 & SME & Software Developer &  Government: Official Website & 14 & \cellcolor{Always!90} Always\\ \hline
        
    P25 & Corporation & Business Analyst &  Customer: Fitness Training App & 4 & \cellcolor{Always!90} Always\\ \hline

    P26 & Corporation & Business Analyst &  Customer: Fitness Training App & 7 & \cellcolor{Always!90} Always\\ \hline

\end{tabular*}

\end{minipage}
\end{center}

\end{sidewaystable}

%% file: Section/Results.tex
\section{Results}\label{re}

In this section, we present the results of the data collection.  For ease of analysis and due to space limitations, we present the results and discussions from both our interviews and survey together.

\subsection{Demographics of Participants}

\sectopic{Interviews.} \textcolor{black}{We interviewed 26 practitioners with an average of eight years of industry experience. All interviewees had at least three years of work experience in various industries. All interviewees worked in different types of IT companies: 14 in large corporations, six in SMEs, and six in startups. One interviewee (P3) was the CEO of an IT company. Five interviewees were located in the United States (P1, P4, P19-21), 15 in China (P2, P3, P5-P14, P17, P18, P25, P26) and five in Australia ( P15, P16, P22-24). Table~\ref{tab:tab1} summarizes the backgrounds of the interviewees.}

\sectopic{Survey.} Our online survey received responses from 323 respondents. Out of these, 75 did not complete the survey, leading to their exclusion. We further filtered out 64 responses from participants who did not align with our job criteria, notably those who lacked a background and were also not working in IT/software fields. This resulted in a final count of 203 valid responses for our analysis.

Figure~\ref{fig:fig1} (a) plots the respondents' experience, with a secondary level breakdown based on their role. Most respondents had 1-4 years of experience (48\%), followed by 5-9 years of experience (24\%), more than ten years (13\%), and less than one year (13\%) in the industry. Figure~\ref{fig:fig1} (b) shows the current roles of respondents. Among the respondents, a large number were software developers (50\%), followed by UI/UX designers and IT project managers (16\% each), user researchers (11\%), and business analysts (6\%). 
We note that while respondents could identify with multiple roles, they were only able to use their primary role in our survey. \textcolor{black}{Figure~\ref{fig:fig1} (c) represents the distribution of roles across varied IT company types. The breakdown was as follows: SMEs (50\%), startups (26\%), and large corporations (24\%).} In our attempt to achieve a globally representative perspective on personas, we sourced responses from a diverse set of geographical regions. We publicized our survey on social media platforms and received responses from three countries -- the USA, Australia, and China. Overall, from our 203 respondents, we had 18 from the United States, 25 from Australia, and the remaining 160 from China.



\input{Section/RQ1}

\input{Section/RQ2}

\input{Section/RQ3}

%% file: Section/RQ1.tex
\subsection{RQ1 Results - Persona Usage}

\begin{figure}[!t]
  \centering
  \includegraphics[width=\linewidth]{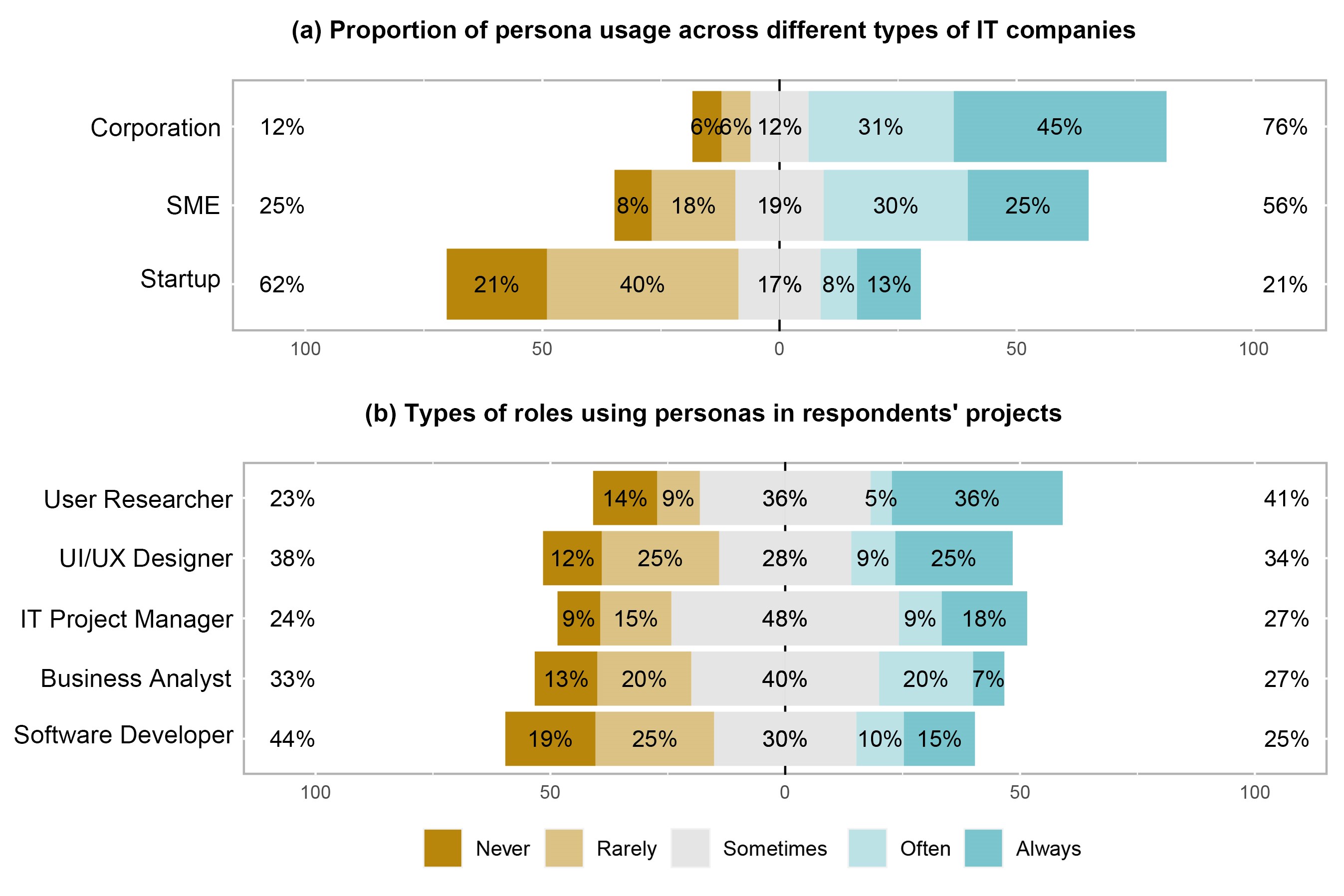}
  \caption{Scaled rates of persona usage in different companies and job roles.}
   \label{fig:Frequency}
\end{figure}

\begin{figure}[!t]
  \centering
  \includegraphics[width=\linewidth]{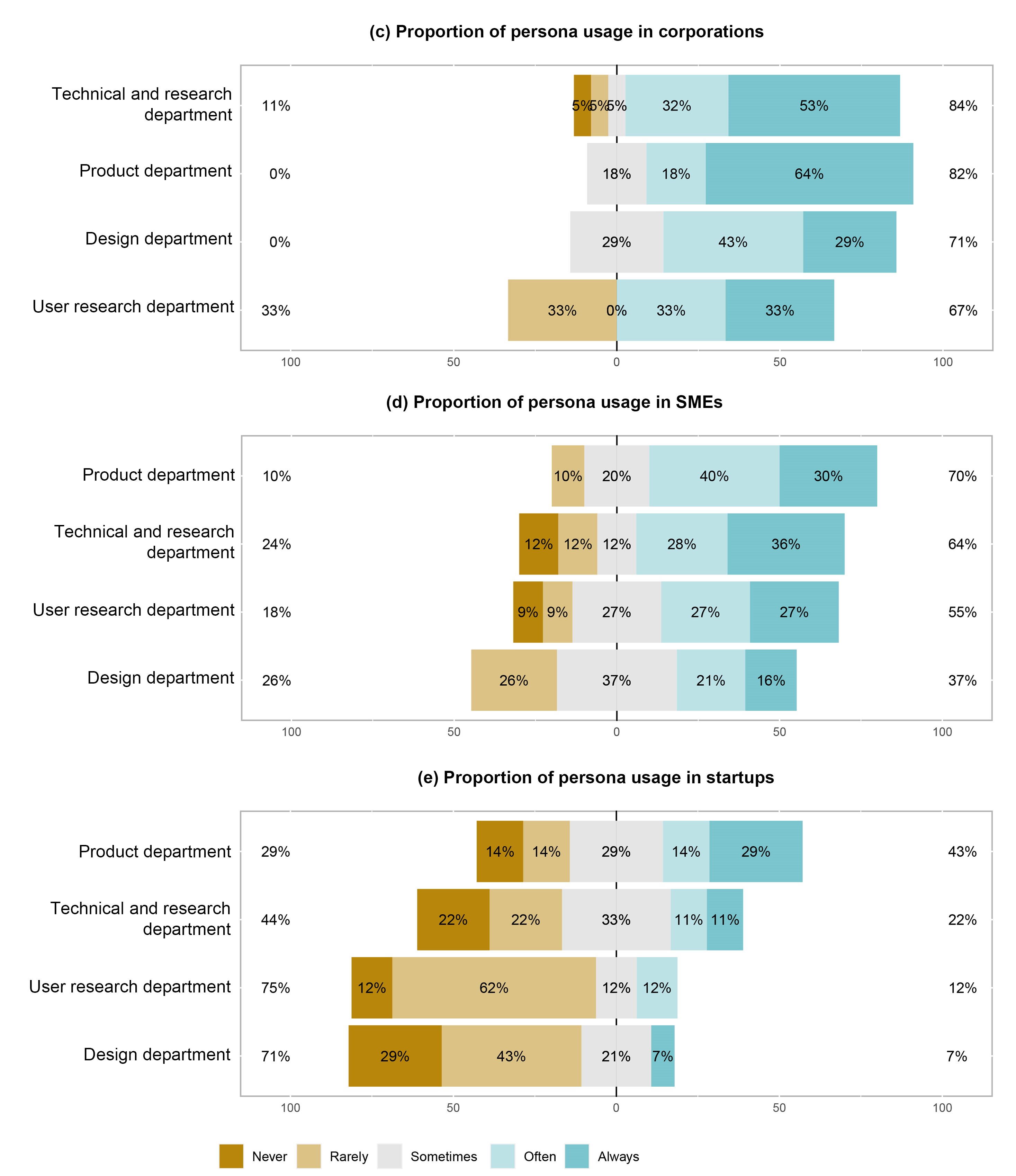}
  \caption{Scaled rates of persona usage in different departments.}
   \label{fig:Frequency1}
\end{figure}

\sectopic{Frequency of persona usage in different-sized companies.} \textcolor{black}{In our survey, we asked a five-point Likert scale question to respondents about the frequency of personas usage in their teams or departments across different types of IT companies, to which around half the respondents (105/203) responded that the personas are used `always' or `often' in their projects, whereas a third of the respondents (64/203) reported that the personas are `rarely' or `never' used. The remaining responded with `sometimes'. \textcolor{black}{Meanwhile, a third of the respondents (74/203) reported that the personas are `rarely' or `never' used by themselves, whereas a third of the respondents (60/203) reported that the persona are `often' or `always' used. The remaining responded with `sometimes'.}} Figure~\ref{fig:Frequency} (a) details this distribution, segmented by company size and respondent roles. 76\% of respondents from larger corporations reported using personas (always/often), whereas only 12\% do not use personas (rarely/never). From SMEs, 55\% respondents use personas while 26\% do not. Startups presented a stark contrast, with only 21\% respondents reported using personas, whereas 61\% reported that they do not. \textcolor{black}{Additionally, Figure \ref{fig:Frequency1}~shows the distribution of persona usage across departments in different types of companies. Overall, large corporations are more popularly to use personas, whereas startups rarely use them in the software development phase. Specifically, 33\% of respondents from user research department and 11\% of respondents from technical and research department in larger corporations reported that they do not use personas (rarely/never), whereas many respondents from all departments in large corporations reported using personas (always/often). 70\% and 43\% respondents from product departments in SMEs and startups respectively reported using personas (always/often). The remaining departments responded fewer using persona than the product departments. It is noteworthy that only 7\% of respondents from startups use personas. } 

In interviews, \textcolor{black}{we had 14 interviewees from corporations and six each from SMEs and startups.} Table~\ref{tab:tab1}'s last column shows the interviewees' responses on persona usage. 
From corporations, nine reported using personas (always/often). Three interviewees from SMEs reported persona usage, and two from startups used personas. Two interviewees (P1 and P5) from corporations considered using personas a valuable tool in software projects for UI/UX designs. P3 provided a plausible explanation of the lack of persona usage in smaller companies and startups: \textit{``Small and medium-sized businesses often lack the resources and expertise to create personas. [...] Large companies have the advantage of having more employees and the ability to hire higher-quality candidates with expertise in user data analysis.''} 
P3 mentioned \textit{``Big corporations have the resources and data to create comprehensive personas, smaller businesses may struggle to do so, as it can be expensive and challenging to generate accurate data.''} In contrast, P17, a project manager at a startup that extensively uses personas, reflected on startups like theirs: \textit{``In the early stages of a startup, personas are used to confirm the ideal target audience and identify their characteristics and requirements. In later stages, personas are used to refine ideas and guide direction and strategy when dealing with different types of users.''} At SMEs, interviewees noted that the use of persona is often linked to project scale.

\tcbset{
    frame code={}
    center title,
    left=0pt,
    right=0pt,
    top=0pt,
    bottom=0pt,
    colback=blue!10,
    colframe=white,
    width=360pt,
    enlarge left by=0mm,
    boxsep=5pt,
    arc=6pt,outer arc=0pt,
    breakable
    }

\begin{tcolorbox}[breakable]
\textbf{Observation \#1.} \textit{Smaller companies, including startups, typically operate with constrained resources and tight project timelines. As a result, they often opt out of creating personas unless a project explicitly demands it. SMEs and larger corporations are more likely to use personas, but the usage may vary based on the project scale.}
\end{tcolorbox}

\textbf{Frequency of persona usage in different roles.} In our survey, we further asked the respondents to clarify if they use personas as a part of their role in their team. Figure~\ref{fig:Frequency} (b) shows the distribution of responses from respondents for their specific roles. User researchers and UI/UX designers are most likely to use personas as part of their roles. There is not much difference in usage among the other three roles, i.e., software developers, project managers, and business analysts. On further clarification of who, according to them, is (or should be) involved in persona building, a large majority of respondents mentioned UI/UX designers, product managers, and user researchers. Some also reported software engineers and data analysts, while some selected not applicable (NA).

\textcolor{black}{In interviews, we had 13 software engineers or developers (incl. P21), four business analysts, five managers (incl. P3), and four UI/UX designers. P1 mentioned their role: }
\textit{``Personas play a role in the design process and support design solutions, resulting in greater customer satisfaction. Product managers may also develop personas, which may have a different format than ours.''} P3 presented his position on why personas are not used in their company: \textit{``Defining user requirements and personas is not the responsibility of software engineers, but of product managers. [...] Product managers may not possess technical backgrounds and may lack the necessary skills in statistics, business analysis, and engineering requirements analysis. This results in an incorrect analysis of the requirements and, therefore, incorrect positioning of personas. When personas are positioned incorrectly, the engineering requirements are misguided, and the project is doomed to fail.''} The role of product manager was mentioned multiple times across interviews as someone who should be responsible for building personas. P3's position gives the idea that the personas building and management are linked to the skills underlying certain roles, on which P2 builds around personas: \textit{``Professional backgrounds such as psychology and social work can greatly benefit our (persona) design process as they bring a deep understanding of users and their needs.''}

\begin{tcolorbox}
\textbf{Observation \#2.} \textit{Persona building is perceived to require a specific skill set, e.g., user data analysis and understanding user needs, and therefore the responsibility is often with roles such as UI/UX designers and product managers.}
\end{tcolorbox}

\begin{figure}[!t]
  \centering
  \includegraphics[width=\linewidth]{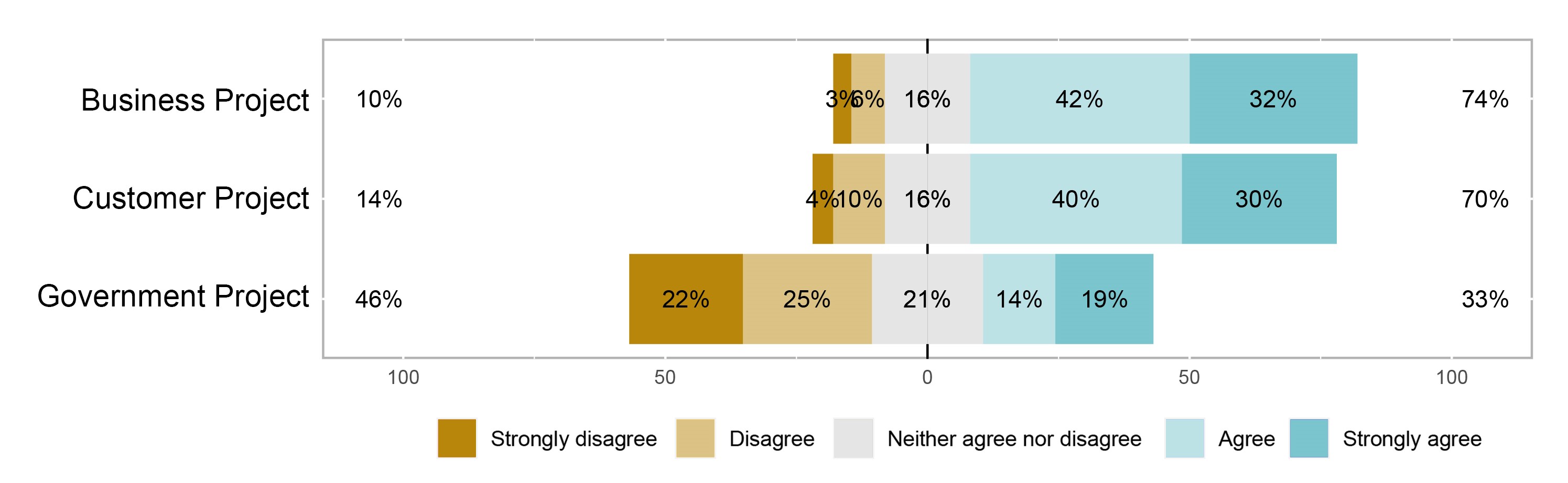}
  \caption{Scaled rates about personas in different types of software projects.}
   \label{fig:fig1.2}
\end{figure}

\textbf{Frequency of persona usage in different types of software projects.}
We mainly investigated three types of software projects in our survey, i.e., projects directly for end users (customers-centric), projects for businesses (e.g., developing systems for other companies or internal CRM systems), or projects for government entities. Respondents reported that personas are used (by saying strongly agree or agree) in customer projects (70\%) and business projects (74\%), but the use of personas in government projects (33\%) is noticeably lower compared to other projects. Figure~\ref{fig:fig1.2} shows the responses for the three software project types.

\textcolor{black}{Two interviewees (P1 and P8) stated that they had experience in developing government projects and mentioned that such projects rarely employ user personas.} 
P8 reported from their previous experiences with government projects: \textit{``In government enterprises, the process often involves a bidding system. The tenders will provide a clear direction for software procurement and outline specific requirements. Therefore, face-to-face interviews are the most effective approach rather than analyzing user needs through personas.''} Regarding business-centric software projects, the interview results varied from the survey, as six out of eight interviewees mentioned that they do not use personas. An overarching reason in most of these interviews was the type of project at hand, i.e., the services for internal clients or employees. P9 said \textit{``The majority of our work is focused on internal operations [...] The users of the internal system can be compared to personas, and we consider their usage habits and needs when designing and developing the system, but we use personas to a lesser extent.''} 

In customer-centric projects, we also observed some variation. P7 noted that others in their company use personas, but they do not as part of their role. \textcolor{black}{P22 and P23 said their projects have a small and clearly-defined user base, e.g., teachers for P22,} and using personas would not be cost-effective. We further noticed (specifically with P4 and P10) that personas are not used in gaming and VR-related projects. Both P4 and P10 noted that they had considered or used personas in the past but consciously chose not to work with personas. This is the case even though user requirements and characteristics are critical in the gaming and VR industries. P4 explained: \textit{``The process of identifying and defining the target audience, or persona, in the video game industry, is unpredictable and constantly evolving.''} P10 explained: \textit{``I currently work in the gaming and VR fields, which differ from conventional software development rules, and there is no mature industry standard for VR. We also hardly ever mention personas in our company.''}

\begin{tcolorbox}
\textbf{Observation \#3.} \textit{Personas are used to a lesser extent in government-related projects than customer or business-centric projects. Within customer-centric projects, the gaming and the VR industry is an exception, where personas are considered less valuable due to rapidly evolving processes.}
\end{tcolorbox}


%% file: Section/RQ2.tex
\subsection{RQ2 Results - Persona Creation Practices}

\textbf{Methods for user data collection and analysis.} We asked respondents in our survey if they use a dedicated data-driven platform for user needs analysis and creating personas. To this, 19\% respondents responded affirmatively. The respondents were further asked about the methods for user need and data analysis, to which $\approx$35\% respondents mentioned using machine learning (ML) or data analytics techniques. The other widely used methods mentioned by the respondents were data visualization techniques and qualitative analysis of user data via questionnaires ($\approx$61\%).

During the interviews, the interviewees shared several different ways of user data collection and analysis in their context. P1 noted \textit{``Our user research department provides statistical tagging based on previous versions of the product, including information such as gender, age, country, and income, which we use to create new personas. \textcolor{black}{An example of the persona creation process is based on the Nielsen method, which consists of ten steps.} We also often use persona templates to streamline the process.''} P2 and P18 both mentioned \textit{``Data mining, observations, (interviews) and questionnaires''} for gathering and analyzing user information. P25 and P26, who work at the same organization, mentioned \textit{``User feedback, complaints and communication with user research and product managers help to understand the users' needs and improve the product through iteration.''}

P13 and P14 (who also work at the same organization) reported on data-driven platforms: \textit{``My company has a data-driven persona platform, as well as employees having an internal persona system''} Similarly, P2 and P20 also commented that they had also used a data-driven persona platform in their previous corporations. They also mentioned that platform development belongs to the work of the user research team. P15 and P2 reported that their companies have focused on developing business processes, data analysis, and visualization platforms. P15 reported an example: \textit{``We are currently developing a results dashboard in a rich form based on survey data. At the moment, data visualizations can be generated, and personal information can be identified some information (human aspects information) [...] like age, gender and so on.''} Additionally, P2 stated: \textit{``Unfortunately, we do not have professional business or data analysts. So, designers focus on requirements analysis using quantitative and qualitative methods.''} In addition, P18 (from startups) mentioned how to use the personas service platform such as Delve.ai~\citep{website22}, and open sources from GitHub~\citep{GitHub}.


\begin{tcolorbox}
\textbf{Observation \#4.} \textit{Various methods are employed for collecting and analyzing user information, from questionnaires and interviews to utilizing feedback from prior product versions, applying ML techniques, and specialized data visualization platforms. While SMEs and larger corporations often deploy internal data-driven systems for persona creation, startups tend to lean towards open-source data-driven platforms.}
\end{tcolorbox}

\begin{figure}[!htb]
  \centering
  \includegraphics[width=\linewidth]{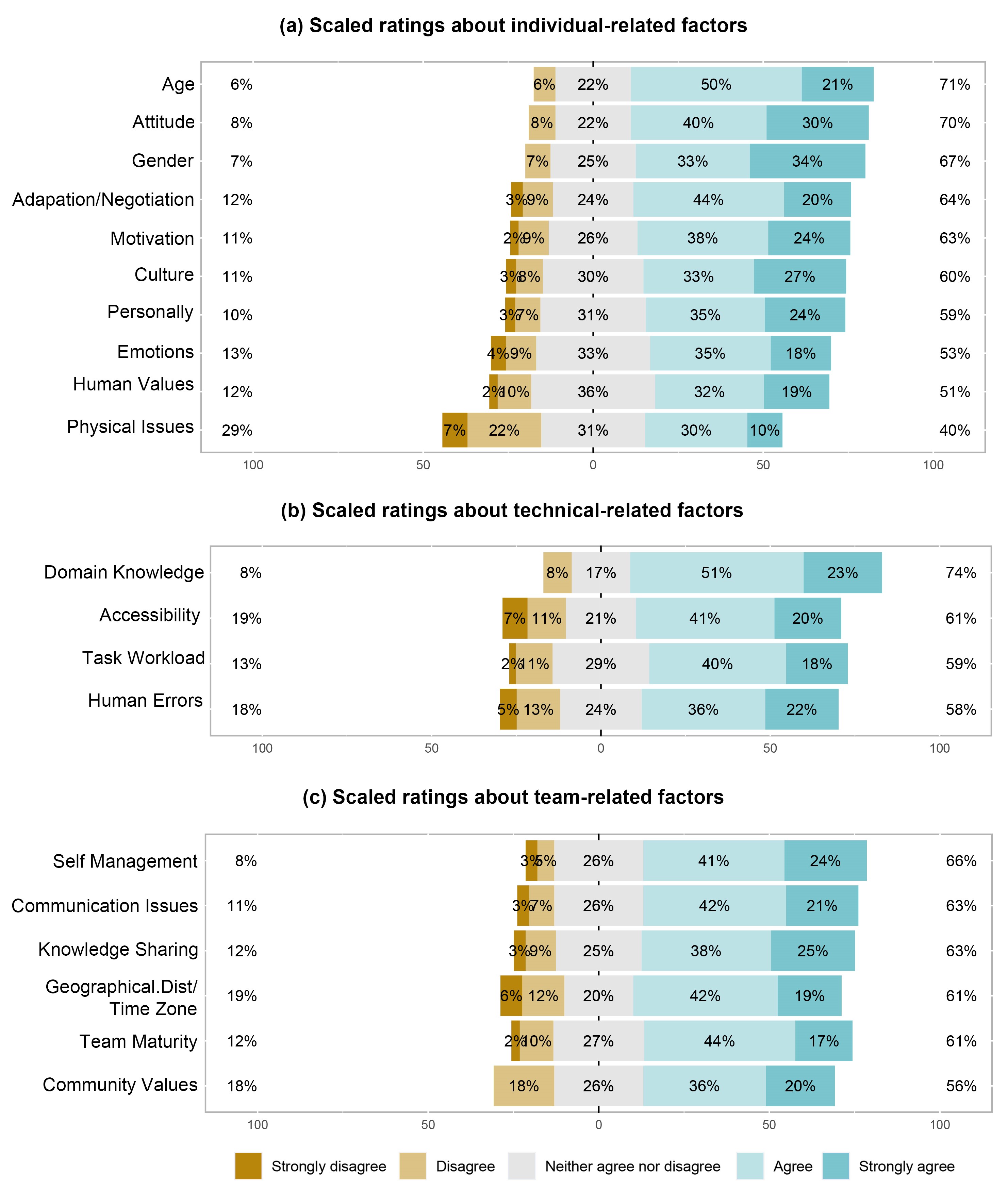}
  \caption{Scaled rates about human aspects in personas.}
   \label{fig:fig3}
\end{figure}
\textbf{Information captured in personas.} We were interested in understanding the user data that is captured in personas, including human-centric factors such as age, gender, and physical issues. When we asked whether respondents considered individual-related, technical-related, and team-related human aspects of requirements in personas, they mostly agreed. Most respondents reported agreeing that demographics were captured through personas. Figure \ref{fig:fig3} (a, b, c) presents the respondents' scaled ratings about their experience and highlights the importance of demographic details in a persona template. \textcolor{black}{Overall, in the creation of personas including human aspects, most human aspects have agreement rates above 50\%, and only physical issues have agreement rates below 50\%.}

Most interviewees considered human aspects beyond basic demographics as important but also mentioned that they do not consider such factors or requirements in general, as they tend to increase the effort and costs associated with personas and RE. Two interviewees (P25 and P26) revealed that when they analyze the requirements of fitness apps, they do not comprehensively consider accessibility requirements or even physical disability issues. We asked whether they consider how special needs users or disabled users use fitness apps. P25 stated: \textit{``Currently our project has ignored these issues. I believe that (physical disability and accessibility issues) are very important, but I do not have the relevant requirements analysis for special users with accessibility issues''.} P26 told us that \textit{``Collecting human aspects data is important in fitness apps, such as time, fitness level, exercise environment, equipment, experience goals, health conditions or restrictions. However, we rarely analyze special user needs or behaviors through user personas.''} Furthermore, P18 explained why special user needs are not considered because \textit{``The proportion of special users is very small. Managers believe that the focus should be on fitness enthusiasts. So, we have hardly ever used user personas for special user groups.''} \textcolor{black}{Only P10 reported that they explicitly consider the interaction of disability groups while eliciting requirements for their VR and gaming design.}

\begin{tcolorbox}
\textbf{Observation \#5.} \textit{Practitioners recognize the significance of incorporating human-centric factors into personas. Despite this acknowledgement, there is a noted lack in their practical application due to concerns about increased costs and efforts, especially for niche user groups like those with physical disabilities. Therefore, unless directly relevant to the core user groups of the systems, the human-centric factors of users are not considered.}
\end{tcolorbox}

In terms of the information captured in personas, P6 noted: \textit{``Personas are not a single entity, but rather a large amount of data that businesses (like ours) collect. Different departments in a company may require different types of data based on the user's characteristics.''}  P6 further reported an example: \textit{``In addition to some common demographic information, we pay more attention to geographic information because we need to collect users' real-time travel information...like country, destination city, time difference and time zone. If we use personas, we can effectively group users with different attributes, providing accurate user groups and requirements for software iterations.''} P4 and P10 mentioned that although they do not use personas, they maintain detailed information on each user, e.g., their demographics, average playtime, start time, time zone, click rate, and favorite font and size. Additionally, P12 stated: \textit{``Important information for an educational app includes users' personal information, knowledge information, team collaboration, self-management, emotional information and so on. These pieces of information can be used to evaluate the user's learning progress over time.''}

\begin{tcolorbox}
\textbf{Observation \#6.} \textit{Companies customize the information captured in personas based on their specific organizational and application needs, ranging from geographic details to user behaviors. The tailored approach ensures a deeper understanding of the user, facilitating more effective user experiences.}
\end{tcolorbox}

%% file: Section/RQ3.tex
\subsection{RQ3. Challenges to using personas}~\label{subsec:RQ3}

\begin{figure}[htp]
  \centering
  \includegraphics[width=\linewidth]{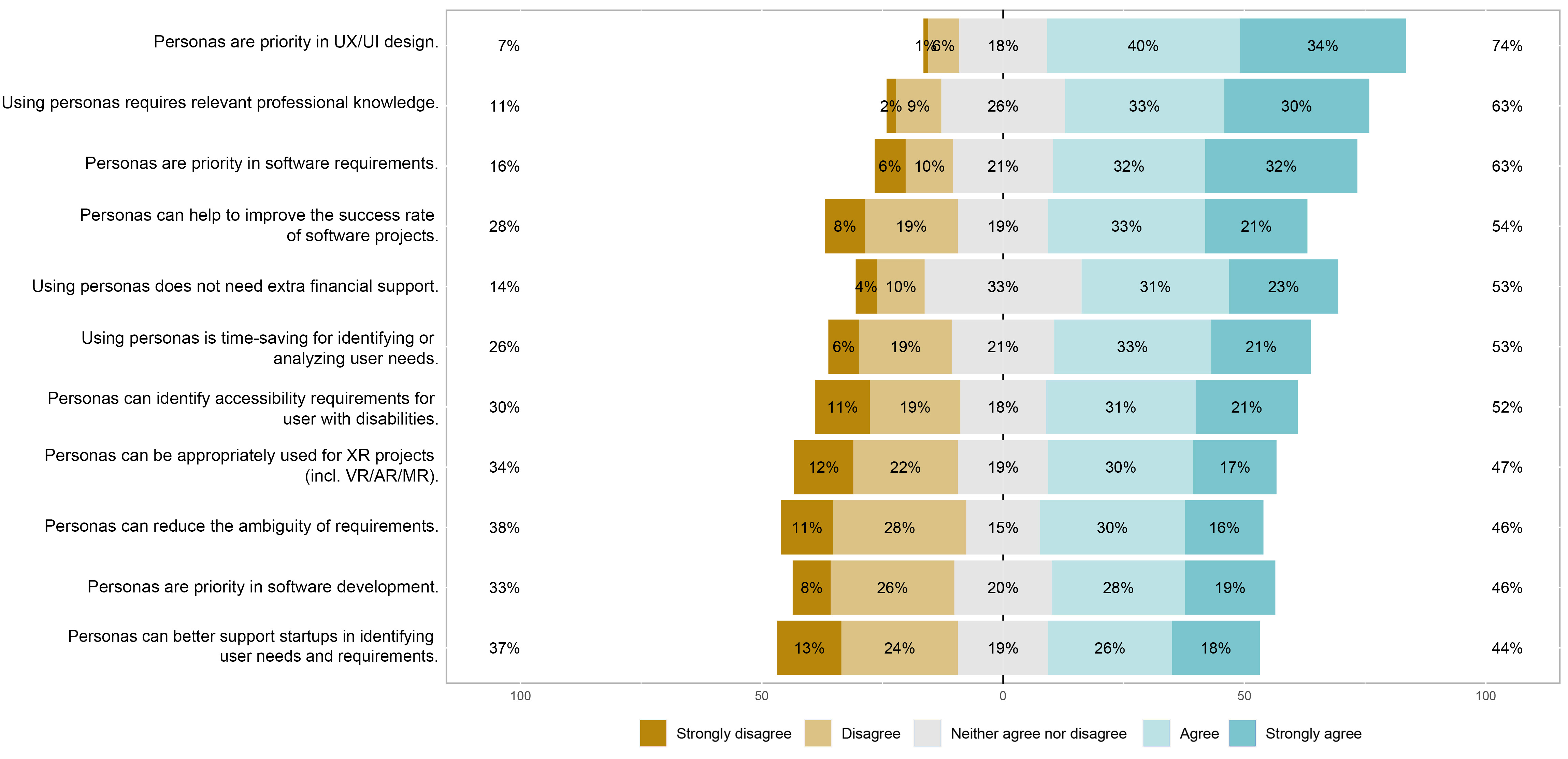}
  \caption{Scaled rated of persona usage by respondent perspectives.}  
   \label{fig:fig4.2}
\end{figure}

\textcolor{black}{In the survey, we asked a series of five-point Likert scale questions about persona usage. Figure \ref{fig:fig4.2} shows the ratings about each question. 74\% and 63\% of respondents reported that personas are a priority approach in UI/UX design and software requirements respectively. However, 46\% of respondents reported that personas are a priority approach in software development. 63\% of the respondents reported that using personas requires relevant professional knowledge, whereas only 11\% of respondents disagreed (strongly disagree/disagree). 54\% of respondents reported that using persona can improve the success of software projects, wheres only 28\% of respondents disagreed. 53\% of respondents reported that using personas does not need additional financial support and can save time in identifying and analyzing user needs, whereas only 12\% and 26\% of repondents disagreed respectively (strongly disagree/disagree). 52\% of respondents reported that personas can help to identify accessibility requirements for user with disabilities, whereas 30\% of respondents disagreed. 47\% of respondents reported that personas can be used for XR (extended reality) projects, whereas 34\% of respondents disagreed (strongly disagree/disagree). 46\% of respondents reported that personas can help reduce the ambiguity of requirements, whereas 38\% of respondents disagreed (strongly disagree/disagree). 44\% of respondents reported that personas help startups identify user needs and requirements, whereas 37\% of respondents disagreed (strongly disagree/disagree).} 

\begin{figure}[htp]
  \centering
  \includegraphics[width=\linewidth]{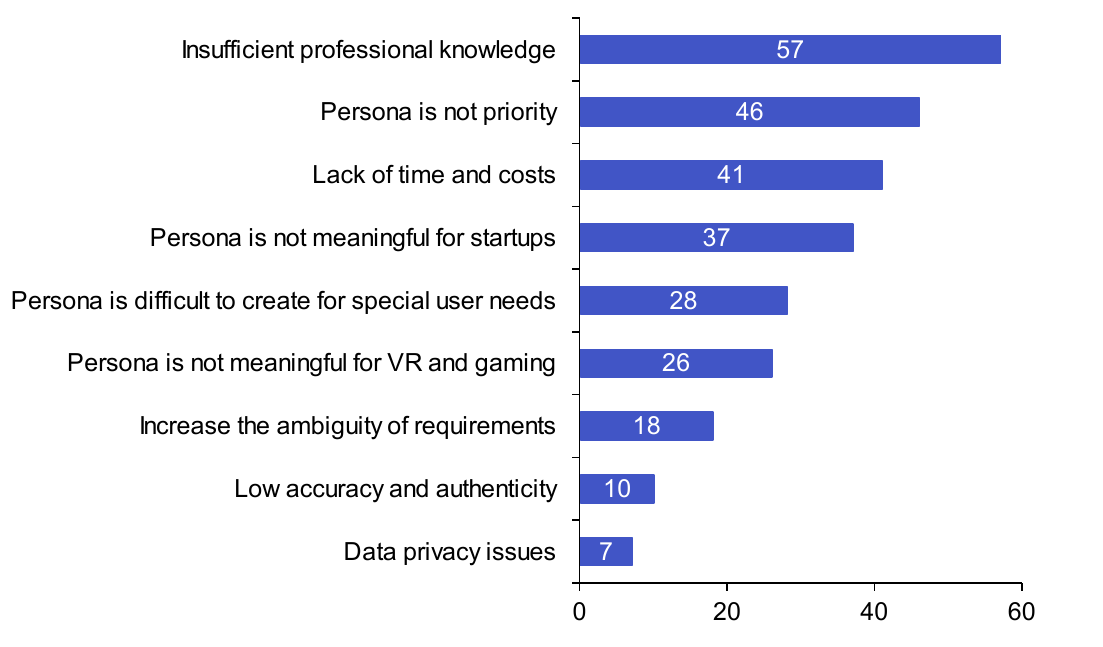}
  \caption{Main challenges concerning personas in industry by interviewees.}  
   \label{fig:fig4.1}
\end{figure}

\textcolor{black}{In interviews, we discuss the key challenges cited by interviewees regarding personas in this section. Figure~\ref{fig:fig4.1} shows interviewees' challenges concerning personas in software projects. }

\textbf{Insufficient professional knowledge.} Insufficient professional knowledge was mentioned 57 times. All interviewees from startups reported that insufficient professional knowledge is one of the key challenges in implementing personas. P3 reported: \textit{``Our team [the designers] primarily collects and analyzes user requirements through interviews. However, designers focus more on aesthetics than scientific data analysis. So, the results are highly subjective.''} Moreover, P2 considered that it is challenging to analyze high-quality results due to the lack of professional background among team members. As discussed in personas usage in different roles in RQ1, interviewees deemed that special skills are required for effectively implementing personas, e.g., the knowledge of statistics, psychology, and user data analysis, which are difficult to find.

\textbf{Time and cost interest.} In total, lack of time and cost was mentioned 41 times. P4 and P12 explained that corporations could invest more resources like time, cost, and personnel and have established and mature frameworks for the requirements process. However, it may not be flexible and require a significant investment of resources for a small-scale project. Most interviewees from SMEs and startups (even if they use personas) confirmed that time and cost are major deciding factors for using personas.

\textbf{Not a priority.} Not a priority was mentioned 56 times. Some interviewees mentioned that personas are not the only method for capturing user data. P16 expressed that specifying user (functional) requirements as documents is seen as a priority rather than ``inefficient'' means such as personas.

\textbf{Right level of abstraction.} Getting the right level of user details is challenging, according to six interviewees. Personas can be inaccurate as they capture only a subset of users or capture too many or too few details. P26 noted: \textit{``Using persona can have drawbacks, limiting our focus to only one segment of the user group and potentially overlooking other important aspects.''} P12 further built on this: \textit{``The challenge is [...] a misaligned persona (can) lead to incorrect decisions about the product's direction and target market.''}

\textbf{Data privacy issues.} Five interviewees mentioned data privacy issues. P15 explained: \textit{``Currently, user data is in a grey area in countries, and legal regulations for user data vary among countries. However, we must ensure that user data is only used securely within our company. I think using user data to create personas can be seen as an invasion of user privacy, but without user data, it is difficult to provide preference services and recommendations for users.''}

\textbf{Maintaining persona databases is challenging.} This challenge was mentioned by five interviewees. P24  elaborated on the challenge: \textit{``It is important to continuously update the persona database. If the data is not up to date, it may not be representative of the actual user demographic.''}

\textbf{Team dynamics.} Three interviewees discussed the challenge of ensuring all stakeholders, including product managers, engineers, and UX designers, are on the same page regarding using and interpreting personas. According to them, if some stakeholders are distant from the process, they do not see the value of creating personas, or they end up clashing with the user requirements or other downstream artifacts.

\textbf{Risk of stereotyping.} Two interviewees mentioned that there is a risk of stereotyping users from a certain background with personas -- \textit{``The information indicated that the product was targeted at 25-35-year-olds and what they were like, which appeared to be fine, but upon further examination, it became clear that 95\% of this persona was incorrect and unreliable (due to stereotyping) [P3].''}



\begin{tcolorbox}
\textbf{Observation 7.} \textit{The challenges (including the ones discussed above) noted by respondents and interviewees:
\begin{itemize}
    \item Professional knowledge (skills) required for personas;
    \item Project pressures, time and costs investment;
    \item Getting the right level of abstraction is difficult, and might lead to missing out on other users;
    \item Data privacy issues;
    \item User personas and requirements might conflict;
    \item Maintaining up-to-date personas is difficult;
    \item Risk of stereotyping;
    \item Team, organizational structure, or the domain (e.g., VR) might not be conducive to persona usage.
\end{itemize}
}
\end{tcolorbox}

%% file: Section/Discussion.tex
\section{Discussion}\label{disc}

In this section, we outline actionable insights from our findings on personas for practitioners and highlight avenues for future research.

\sectopic{Implications for Practitioners.} 

\begin{itemize}[leftmargin=*]
\item \emph{Focused training on Personas.} Our results highlight inconsistent use of personas across different job roles. Specifically, software developers tend to focus less on personas, while job roles more closely related to user needs tend to use personas more frequently. 
\emph{[Takeaway]} This disparity in persona usage emphasizes the need for improved education and training on the benefits of using personas and how to integrate them effectively into the software development lifecycle~\citep{arora2023persona}.

\item \emph{Engage Software Engineers in Persona Development.} As suggested by P5 in their interview, encouraging software engineers to participate in RE-related discussions, including persona creation, can lead to a deeper understanding of user needs.\newline 
\emph{[Takeaway]} Regularly schedule collaborative sessions between software engineers/developers and the requirements analysts and product owners to discuss user needs and co-create personas.
\item \emph{Understand the limitations.} Our study highlights the potential limitations of using personas (discussed in Section~\ref{subsec:RQ3}). For example, some startup practitioners view personas as unnecessary due to lacking resources, data, or professional knowledge. Additionally, subjective personas without empirical data may negatively impact software development. \newline 
\emph{[Takeaway]} Before diving into persona creation, practitioners should understand its potential limitations, especially in resource-constrained environments, and ensure that they are backed by empirical data.

\item \emph{Data-driven persona platforms may help.} Several practitioners reported that their organizations successfully use third-party or internal data-driven persona platforms. 
\newline 
\emph{[Takeaway]} Organizations should explore and consider investing in data-driven platforms to streamline and automate the persona creation process, to cater to diverse user requirements.

\item \emph{Ensure accurate persona content coverage.} Our study suggests that the content of personas should accurately reflect user needs and behaviors. This could include exploring how human factors and user requirements map onto personas and investigating effective methods for constructing and validating personas. 
\newline 
\emph{[Takeaway]} Regularly validate and update personas. Consider workshops or feedback sessions to map human factors and user requirements onto personas effectively, to improve software products' quality and user experience.
\end{itemize}

\sectopic{Implications for researchers.} Below, we present potential research avenues and recommendations derived from our findings.
\begin{itemize}[leftmargin=*]
    \item \emph{Usage of personas in resource-constrained development environments.} Many participants noted that they do not implement personas due to resource shortages. The possible research directions are related to exploring how practitioners can create and use personas effectively despite limited resources. For instance, can RE researchers leverage natural language processing (NLP) solutions, such as large language models (GPT Model \citep{zhang2023personagen})? Alternatively, what other lightweight options can practitioners use in lieu of personas in resource-constrained environments?
    
    \item \emph{Investigate the various impacts of data-driven personas in the industry.} The open research questions related to data-driven platforms, based on our findings are: How effective are data-driven persona platforms in different contexts and for different software products? Investigate the impact of data-driven persona platforms on the quality and user experience of software products that use such platforms. In light of the privacy concerns raised by our participants, what are the ethical and privacy implications of using data-driven persona platforms?

    \item \emph{Consideration of human aspects in personas.} As discussed earlier in the results of our study, many human aspects, e.g., special physical needs, were largely ignored in personas. Can we develop guidelines and best practices for building personas inclusive of the diverse user bases in different systems?
    
    
    \item \emph{Usage of personas in constantly involving application domains.} Our study found that personas are `rarely' or `never' used in VR and gaming software projects, as these fields are constantly evolving. One possible reason could be that the user experience and interaction in VR and gaming environments are perceived as different from those in traditional software applications. If this is the case, then a possible research direction is to study these perceptions of user experience and investigate the user-centered RE methods for VR and gaming projects.

 \textcolor{black}{   \item \emph{Usage of personas in varying types of software projects and IT companies.} Our study found that personas are `rarely' or `never' used in startups, and government software projects. Our study reveals the frequency of persona usage across different types of projects and IT companies. In the future, we should focus on employing personas to enhance practices in various types of software projects, such as government-based websites, applications, extended reality (XR), data visualization, culture exhibition projects and so on. We also should focus on how personas can effectively help in identifying and analyzing user needs and requirements for startups, or small/medium sized IT companies, evening focusing the differences across different countries and regions. }
    
\end{itemize}

%% file: Section/Limitation.tex
\section{Threats to Validity and Limitations}\label{lim}

Various factors were potential candidates as threats to the validity of our qualitative study, e.g., research bias, which involves selectively recording information and confirming personal biases, as well as inaccurate responses~\cite{Kovalenko20, onwuegbuzie2007, johnson1997}. In the following, we outline our strategies for mitigating potential threats to the validity. 

During the interview, there was a possibility that we may have inadvertently influenced participants to provide researchers with desirable answers. To mitigate researcher bias, we employed several strategies. Firstly, we recruited participants from various fields and had many interviewees. Secondly, we arranged for multiple researchers to participate in the interviews and collect and analyze the data. Thirdly, the researchers avoided in-depth discussions on socially acceptable answers while providing important information about the study's purpose and topic. Another potential threat is participant bias, where some participants may not provide accurate answers due to their experience or biases. To avoid this, we ensured participants and their company names remain anonymous~\cite{furnham1986}. 

To mitigate respondents' bias during the survey, we employed several strategies. Firstly, we asked interviewees, their personal industrial networks, and the professional research team's social media platform to distribute our online survey. \textcolor{black}{Secondly, we invited three experts from the fields of requirements engineering and user-centered design in the survey design process, and two industry professionals (more than five years of experiences in software requirements) to evaluate our research questions with us to ensure that respondents could easily understand the questions' meaning.} Thirdly, we conducted an initial survey to assess the quality of respondents' answers before the final evaluation, which was not included in this study. Fourthly, the survey went through four rounds of evaluation and modification before its final version. In addition, we added an introduction on the initial page to provide information about our study.

To ensure data quality, we sourced practitioners from diverse sources: job roles, software projects and IT companies. We excluded data that was ambiguous, incomplete, or where respondents were unfamiliar with personas, or repeatedly selected the same answers. Finally, to ensure the validity of our findings, we employed triangulation methods, enhancing result accuracy and reliability. Upon analyzing the interviews, we noted that the majority of our survey findings aligned with the interview outcomes.

While our study offers valuable insights and substantial contributions to the understanding of persona usage in SE, it is not without its limitations. Our participant pool predominantly comprises practitioners from Australia, China, and the United States, potentially not capturing a full global perspective. While the insights gleaned are valuable, we recognize the need for a broader global perspective and thus intend to expand our research scope to capture the viewpoints of practitioners from other underrepresented areas. \textcolor{black}{Furthermore, some interviewees and respondents have never used personas represents a potential threat. To mitigate this threat, we required all interviewees to understand the concept of personas, which may provide valuable insights on why personas are not used in different types of IT company. We also added a reminder at the beginning of the survey to prevent respondents who are unfamiliar with personas from answering the questions. We removed all responses that were similar answers, and set up automatic function on the online platform to identify invalid responses. Additionally, we removed job roles that are unrelated to software and user requirements. } 

%% file: Section/Conclusion.tex
\section{Conclusion}\label{conc}

This paper explores the use of personas across different industry sectors. Our study indicates that while SMEs and corporations frequently use personas, startups rarely do so. Our interviews revealed five prominent job roles that employ personas: UI/UX designers, IT project managers, software developers, business analysts, and user researchers. We also observed that persona usage varies considerably depending on the type of software project. Specifically, government-related projects utilize personas the least, whereas customer and business software projects often employ them. Moreover, we discerned that IT companies and specific job roles demonstrate diverse preferences, methodologies, and attitudes towards persona usage. Notably, IT firms face challenges when utilizing personas, such as limited professional expertise, resource constraints, and data privacy concerns. Intriguingly, our study pinpointed several crucial human aspects that are often overlooked in personas. From our insights, we offer recommendations for both practitioners and researchers. In the future, we aspire to delve deeper into the potential of data-driven platforms and cutting-edge technologies, like VR, to navigate these challenges and cover a wider participant pool.